\documentclass[twocolumn]{aastex631}
\usepackage{xspace}
\usepackage{natbib}
\usepackage{amsmath,upgreek}

\newcommand{\Ha}{H$\alpha$\xspace}
\newcommand{\Lacc}{\ensuremath{L_{\mathrm{acc}}}\xspace}
\newcommand{\LHa}{\ensuremath{L_{\mathrm{H}\alpha}}\xspace}
\newcommand{\MMd}{\ensuremath{M\dot M}\xspace}
\newcommand{\MdtoM}{\ensuremath{\dot M\textrm{--}M}\xspace}
\newcommand{\LHatoLacc}{\ensuremath{L_{\mathrm{H}\alpha}\textrm{--}L_{\mathrm{acc}}}\xspace}

\newcommand{\MMdunits}{\ensuremath{M_\mathrm{J}^2\,\rm{yr}^{-1}}\xspace}

\defcitealias{Betti2023}{B23}
\defcitealias{Stamatellos2015}{SH15}
\defcitealias{Alcala2017}{A17}
\defcitealias{Aoyama2021}{A21}

\newcommand{\complst}{7.47\xspace}
\newcommand{\complpl}{5.46\xspace}

\newcommand{\ppratestelluncert}{$0.16^{+0.19}_{-0.10}$\xspace}
\newcommand{\pprateplanuncert}{$0.22^{+0.26}_{-0.14}$\xspace}
\newcommand{\ingapratestelluncert}{$0.29^{+0.24}_{-0.15}$\xspace}
\newcommand{\ingaprateplanuncert}{$0.40^{+0.33}_{-0.21}$\xspace}

\newcommand{\mediancompratestell}{0.42\xspace}
\newcommand{\mediancomprateplan}{0.57\xspace}
\newcommand{\medianingapstel}{0.29\xspace}
\newcommand{\medianingapplan}{0.40\xspace}
\newcommand{\medianppstel}{0.16\xspace}
\newcommand{\medianppplan}{0.22\xspace}

\newcommand{\MLcompstel}{0.34\xspace}
\newcommand{\MLcompplan}{0.46\xspace}
\newcommand{\MLingapstel}{0.20\xspace}
\newcommand{\MLingapplan}{0.28\xspace}
\newcommand{\MLppstel}{0.07\xspace}
\newcommand{\MLppplan}{0.09\xspace}

\newcommand{\CIcompstel}{$0.23-0.70$\xspace}
\newcommand{\CIcompplan}{$0.31-0.93$\xspace}

\begin{document}

\title{Occurrence rates of accreting companions from a new method for computing emission-line survey sensitivity: application to the H$\alpha$ Giant Accreting Protoplanet Survey}

\correspondingauthor{Cailin Plunkett}
\email{caplunk@mit.edu}

\author[0000-0002-1144-6708]{Cailin Plunkett}
\affiliation{Department of Physics and Kavli Institute for Astrophysics and Space Research, Massachusetts Institute of Technology, 77 Massachusetts Ave, Cambridge, MA 02139, USA}
\affiliation{LIGO Laboratory, Massachusetts Institute of Technology, 185 Albany St, Cambridge, MA 02139, USA}
\affiliation{Department of Physics \& Astronomy, Amherst College, 25  East Drive, Amherst, MA 01002, USA}

\author[0000-0002-7821-0695]{Katherine B. Follette}
\affiliation{Department of Physics \& Astronomy, Amherst College, 25  East Drive, Amherst, MA 01002, USA}

\author[0000-0002-2919-7500]{Gabriel-Dominique Marleau}
\affiliation{Max-Planck-Institut für Astronomie, Königstuhl 17, D-69117 Heidelberg, Germany}
\affiliation{Fakultät für Physik, Universität Duisburg-Essen, Lotharstraße 1, D-47057 Duisburg, Germany}
\affiliation{Institut für Astronomie und Astrophysik, Universität Tübingen, Auf der Morgenstelle 10, D-72076 Tübingen, Germany}
\affiliation{Physikalisches Institut, Universit\"{a}t Bern, Gesellschaftsstr.~6, 3012 Bern, Switzerland}

\author[0000-0001-6975-9056]{Eric L. Nielsen}
\affiliation{Department of Astronomy, New Mexico State University, Las Cruces, NM 88003}

\begin{abstract}

A key scientific goal of exoplanet surveys is to characterize the underlying population of planets in the local galaxy. In particular, the properties of accreting \textit{proto}planets can inform the rates and physical processes of planet formation. We develop a novel method to compute sensitivity to protoplanets in emission-line direct-imaging surveys, enabling estimates of protoplanet population properties under various planetary accretion and formation theories. In this work, we specialize to the case of \Ha and investigate three formation models governing the planetary-mass-to-mass-accretion-rate power law, and two accretion models that describe the scaling between total accretion luminosity and observable \Ha line luminosity. We apply our method to the results of the Magellan Giant Accreting Protoplanet Survey to place the first constraints on accreting companion occurrence rates in systems with transitional circumstellar disks. We compute the posterior probability for transitional disk systems to host an accreting companion ($-7\leq\log \MMd /(\MMdunits)\leq-3$) within 2~'' ($\sim200$~au). Using our two accretion models, we find consistent protoplanet rates, with median and 68\% credible intervals of \ppratestelluncert and \pprateplanuncert accreting companions per star, respectively. Our technique enables studying protoplanet populations under flexible assumptions about planet formation. This formalism provides the statistical underpinning necessary for protoplanet surveys to discriminate among formation and accretion theories for planets and brown dwarfs.
\end{abstract}

\keywords{Accretion (14), Brown dwarfs (185), Direct imaging (387), Exoplanet formation (492), Exoplanet detection methods (489), H alpha photometry (691)}

\section{Introduction} \label{sec:intro}

Theoretical and observational studies over the past half century have constrained many aspects of planet formation, drawing from simulations as well as statistical studies of planet populations~\citep[e.g.,][]{Toomre1964,Safronov1972,Reipurth2001,Stamatellos2009,Lambrechts2012,Forgan2013,Stamatellos2015,Vigan2017, Fernandes2019, Bowler2020, DoO2023, Squicciarini2025}. In particular, large-scale direct imaging campaigns such as the Gemini Planet Imager Exoplanet Survey~\citep{Nielsen2019} and the SPHERE Infrared Exoplanet survey~\citep{Vigan2021,Chomez2025} have demonstrated the relatively low occurrence of giant planets at wide separations ($\gtrsim10$~au). However, the confirmed existence of some wide-separation giant planets challenges the classic model of planet formation by core accretion.

Protoplanets that are actively accreting matter provide direct windows into the early stages of planet formation, making them ideal targets for discriminating among formation theories.  Young stars with ``transitional'' disks, which host large central cavities thought to be cleared by forming planets~\citep{Lin1993,Dodson2011,Price2018,Close2020}, have proven fruitful targets for protoplanet imaging. Recent surveys of these have revealed several protoplanets and candidates \citep[e.g.][]{Zurlo2020, Haffert2021,Huelamo2022,Follette2023}. Dedicated second-generation protoplanet imaging surveys \citep[][]{Close2020SPIE,Zhou2021prop,Chilcote2022} will continue to increase their numbers.

\subsection{The difficulty of protoplanet imaging}

Circumstellar environments present obstacles for detection and validation of protoplanet candidates. Complex disk morphologies impede ready separation of disk and planet signals~\citep{Follette2017, Currie2019}. As such, of all candidates reported within circumstellar disk gaps, only PDS~70~b and PDS~70~c~\citep{Keppler2018,Haffert2019,Wang2021,Zhou2021} are considered unambiguous protoplanets, while the natures of many others remain debated~\citep[e.g., but not limited to,][]{Kraus2012,Reggiani2014,Sallum2015,Follette2017,Rameau2017,Currie2019,Gratton2019,Currie2022,Zhou2022,Zhou2023,Hammond2023,Biddle2024,Currie2024}.

On the detection side, most protoplanet candidates have been sought and detected only in narrowband accretion-tracing spectral emission lines where protoplanets' brightnesses are enhanced relative to those of their host stars (and the photospheric contribution is assumed to be negligible). \Ha, which is both bright and accessible from the ground, is a common choice. However, the paucity of confirmed protoplanets has spurred research into why they may be difficult to detect in \Ha. \citet{Brittain2020} theorized planets undergo episodic accretion, thus making it less likely to capture a planet during an accreting phrase. Indeed, the \Ha fluxes of PDS~70~b and~c have been observed to vary by factors of $\sim5$ and 2 on a $\sim1$ year timescale~(\citealp{Close2025}, \citealp{Zhou_2025_accvar} independently find compatible results); AB~Aur~b has similarly shown \Ha variability $\sim20$ times that of its host star~\citep{Bowler2025}. Planetary accretion mechanisms may also result in weak \Ha production efficiency, particularly at low accretion rates~\citep{Thanathibodee2019,Aoyama2021}. The lack of detections has complicated efforts to resolve the tensions in planet formation models.

While not strictly protoplanets, numerous wide-separation substellar objects have been found to be accreting, which also oppose classical companion formation mechanisms. For example, the low-mass companions DH~Tau~b, GQ~Lup~b, GSC~06214-00210~b, and Delorme 1 (AB)b have shown excess emission in one or more of: \Ha, He I, Paschen $\beta$, Paschen $\gamma$, and the NUV/optical continuum, suggesting strong ongoing accretion~\citep{Zhou2014, Eriksson2020, Stolker2021, Betti2022}. Given their separations, the existence of these wide companions implies formation in a massive extended disk around the host star, a star-like binary formation mechanism, or dynamical evolution that drives the companions outward after formation. Their accretion rates are higher than standard \MdtoM relations predict, which~\citet{Zhou2014} interpret as evidence for formation by gravitational instability in a massive protostellar disk. Moreover, their line profiles imply a magnetospheric or ``star-like" accretion mechanism~\citep{Demars2023} (see Sections~\ref{ssec:formpar}--\ref{ssec:accpar} for further definition and discussion of these mechanisms). Also of interest is the existence of free-floating, low-mass accreting objects, such as 2M~1115~\citep{Theissen2018, Viswanath2024}, OTS~44~\citep{Joergens2013}, and TWA~27~B~\citep{Luhman2023,Marleau2024}, whose formation and accretion properties may differ from bound objects. Such systems provide useful analogs for planet formation environments and enable testing formation mechanisms across the mass spectrum.

\subsection{Protoplanet survey selection effects}

The difficulties of imaging substellar objects mean survey results are biased toward the brightest, most easily detectable objects: those most massive and strongly accreting. To understand planet formation in full, we must conduct population inference based on survey results. This necessitates rigorous accounting for selection effects. Specifically, population studies must quantify the \textit{completeness to planets}, or selection function: the proportion of planets that could have been detected around survey stars as a function of the planets' physical properties. A method to estimate completeness for fully-formed ($\gtrsim 10$ Myr) directly-imaged planets, based on contrast limits and models for planet orbits and luminosities, has been developed over the past $\sim15$ years~(e.g. \citealt{Biller2007, Nielsen2008,Wahhaj2013a,Bowler2016,Nielsen2019, Chomez2025}). In advance of next-generation protoplanet imaging surveys, it is critical to formulate statistical methods to estimate the selection function for \textit{proto}planets, thereby facilitating constraints on the rates and processes of planet formation.

The critical difference between fully-formed planets and protoplanets for computing selection effects is in estimating the intrinsic luminosity from mass. This is because protoplanet luminosity includes contributions from the protoplanetary photosphere, infalling accreting material, and planetary surface shock. These must be disentangled to infer the underlying properties of the object. Moreover, as most protoplanet surveys have imaged in narrowband lines like \Ha, computing detection probabilities (and therefore population properties) for protoplanets requires modeling how protoplanets' masses and accretion rates manifest in the observed luminosities of accretion-tracing spectral lines. Section~\ref{sec:models} details how this is a function of (1) the formation condition of the object and its disk; and (2) the mechanism of accretion onto the object.

In this article, we present a method for computing survey sensitivity to accreting protoplanets under flexible assumptions about formation and accretion, enabling the first statistical study of protoplanet population properties. Section~\ref{sec:models} outlines the models for planet formation and accretion we apply to our dataset. In Section~\ref{sec:data}, we describe the Magellan Giant Accreting Protoplanet Survey~\citep[GAPlanetS,][]{Follette2023}, the survey used for our analysis. Section~\ref{sec:methods} describes our Monte Carlo simulation technique for computing completeness. We present a general framework, then implement the astrophysical models of Section~\ref{sec:models}, and apply the technique to GAPlanetS data. We report completeness estimates in Section~\ref{sec:completeness} and constraints on companion occurrence rates in Section~\ref{sec:constraints}, before concluding in Section~\ref{sec:conclusion}.

\section{Planet formation and accretion}\label{sec:models}

Computing the probability of detecting a planet of given physical properties requires converting its physical parameters to those observed, then comparing its observable characteristics to a threshold for detection. In protoplanet high-contrast imaging, the measurable parameters are: (1)~the projected angular separation (in milliarcseconds); and (2)~planet-to-star line luminosity ratio (contrast; here, we consider imaging at \Ha). The physical properties of interest, on the other hand, are (1)~the semimajor axis $a$; (2)~the mass $M$; and (3)~the mass accretion rate $\dot M$. It is worth noting that estimating selection effects from models and interpreting observational data require opposite chains of models. Computing completeness requires forward modeling from physical parameters to observable ones, while interpreting observations requires statistically linking the data to the physical properties. For protoplanets, the mass maps to the observable (\LHa, assuming \LHa of the star is known) according to the following schema:
\begin{align}\label{eq:mapping}
    M \to \MMd \to \Lacc \to \LHa.
\end{align}
The relevant equation for each step of this mapping depends on the processes that formed the object and on its accretion mechanism. The formation condition determines the mass of the circumsecondary disk surrounding an accreting object---i.e., the amount of matter available for accretion---which subsequently controls the expected accretion rate of a companion of a given mass. The accretion mechanism governs the fraction of total accretion emission that escapes in the observed spectral line.

There is at least one additional aspect of converting protoplanet mass $M$ to observables, which is correcting for extinction local to the planet. Disk or circumplanetary material can obscure accretion columns and shock zones, decreasing the observed line flux. The details are highly simulation dependent, but \citet{Szulagyi2020} showed dust extinction can quench \Ha emission from planets $\lesssim \, 10 M_\mathrm{J}$. Local extinction has been employed to interpret non-detections in systems such as PDS~70 and AB~Aur~\citep{Aoyama2019,Cugno2021, Uyama2021, Biddle2024}, and strengthened the case for observations in additional line tracers less subject to extinction.

In this work, we assume the extinction toward the planet is the same as toward the primary star. Since the target planets lie in well-cleared transitional disk gaps, we expect the interstellar extinction to be dominant. Ideally, one would account explicitly for local extinction, both from the circumplanetary disk and accretion region itself. While recent progress has been made on modeling the dust properties around protoplanets~\citep{Krapp2022,Shibaike2024,Schulik2025}, there do not yet exist methods to robustly estimate extinction for forming planetary systems, particularly as the unknown system geometry may impact the line-of-sight extinction. Not accounting for circumplanetary extinction is a caveat of this work. We discuss its potential impact in Section~\ref{sec:conclusion}.

\begin{figure*}[!htb]
    \centering
    \includegraphics[width=0.9\textwidth]{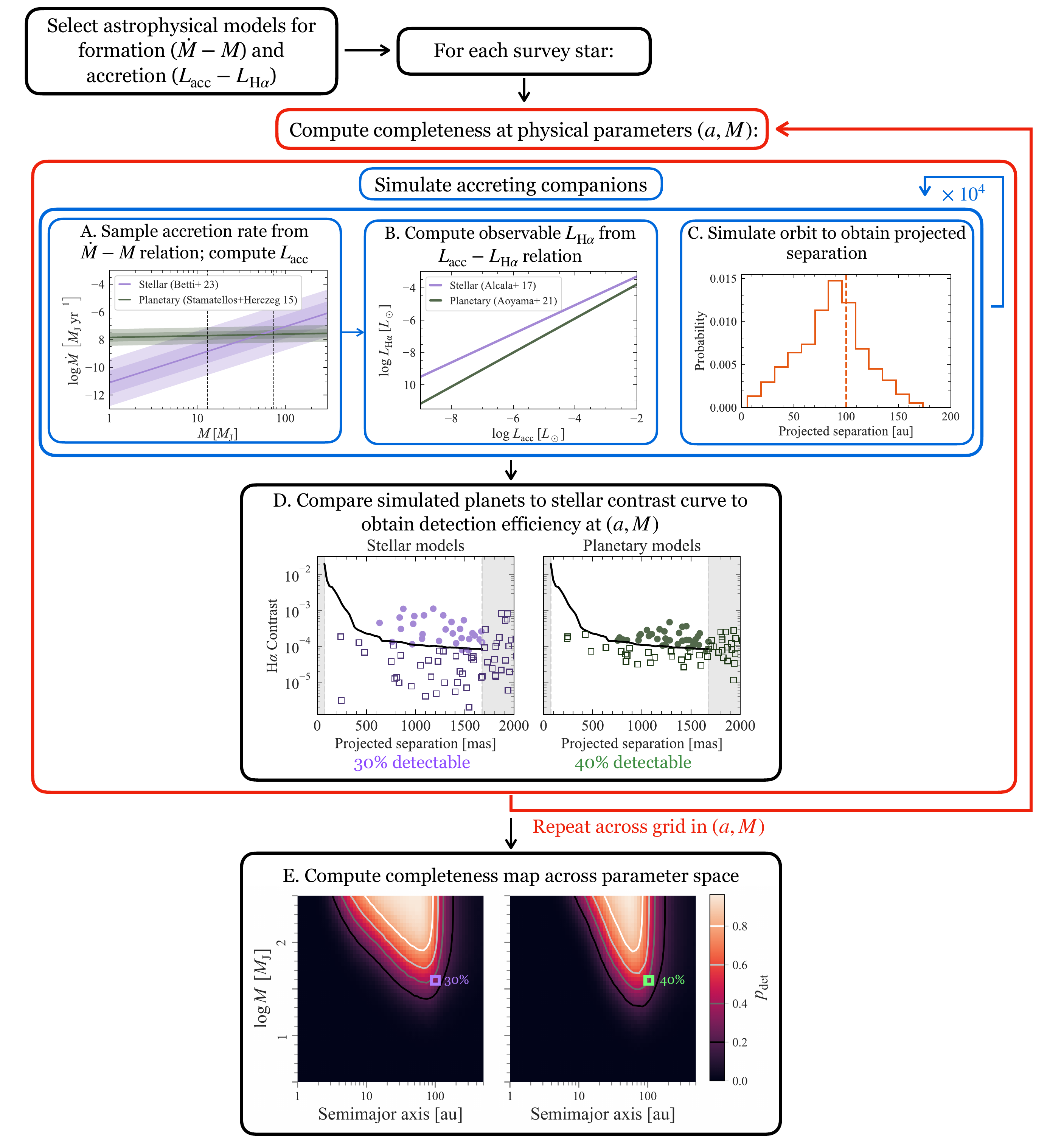}
    \caption{Workflow for estimating direct-imaging survey completeness to accreting companions. First, select models for \MdtoM and \LHatoLacc, corresponding to assumptions about formation and accretion. Estimate the completeness for each survey star with the following Monte Carlo procedure. At each set of physical parameters $(a,M)$, simulate $10^4$ accreting companions.
    (A) Sample mass accretion rates $\dot M$ using an assumed model for the \MdtoM scaling. We take the empirical fit of~\citet{Betti2023} as a more ``stellar'' scenario (purple) and the result of~\citet{Stamatellos2015} as a more ``planetary'' model (green; for both, see Section~\ref{ssec:formpar}). The colored bands are the 1$\sigma$ and 2$\sigma$ uncertainties. We convert each $\MMd$ to $\Lacc$ (Section~\ref{ssec:gravinfall}). (B) Compute the observable $\LHa$ based on a model for the accretion scaling. This is either (see Section~\ref{ssec:accpar}): an empirical, stellar magnetospheric scaling (purple; \citealp{Alcala2017}) or a theoretical planetary relation (green; \citealp{Aoyama2021}). (C) Obtain the projected separation distribution by sampling companion orbital parameters (Section~\ref{ssec:orbits}); the histogram is an example for an object at $a=100$~au with eccentricities following~\citet{Nielsen2019}. (D) Next, compare the simulated companions to the star's contrast curve(s) to estimate the detectable fraction. Panel~D shows two examples of 100 brown dwarfs with $(a,M) = (100\,\mathrm{au},\, 40\,M_\mathrm{J})$ around TW~Hya; the left image shows companions simulated using both ``stellar" models, while the right image uses both ``planetary" models. The light circles are recovered synthetic companions, while the dark squares are undetected. (E) Repeat across a grid in $(a.M)$ for each combination of models to obtain the entire completeness map. Take the average across the survey stars. Repeat for each combination of astrophysical models of interest.}
    \label{fig:flowchart}
\end{figure*}

The astrophysical models we employ at each step are detailed in the remainder of this section. For each, we compare one ``stellar" model, derived from empirical data, and one theoretically-derived ``planetary" model. We then compute completeness following a Monte Carlo procedure that implements these steps, shown schematically in Figure~\ref{fig:flowchart}. The simulation details are in Section~\ref{sec:methods}.

More specifically, Section~\ref{ssec:mmdot_obs} summarizes current observational evidence for a power-law relation between object mass and mass accretion rate, the first step in Equation~\ref{eq:mapping}. In Section~\ref{ssec:formpar}, we describe how the slope and spread of this relation reflects the formation pathway of the planets and their disks. We then introduce two formation models we choose for this work. The relation between $\MMd$ and accretion luminosity \Lacc (the second step of Equation~\ref{eq:mapping}) follows classic assumptions for gravitational infall (Section~\ref{ssec:gravinfall}). Section~\ref{ssec:accpar} describes the models we adopt in this work for the proportion of total accretion luminosity \Lacc that escapes in \Ha (the third step of Equation~\ref{eq:mapping}). This fraction depends on the physical properties of the accretion flow. Finally, in Section~\ref{ssec:orbits} we describe how a companion's projected separation is a function of the system's orbital parameters, a computation that closely follows established methods.

With this framework, we can compute observable characteristics of protoplanets as a function of semimajor axis $a$ and mass $M$ under any combination of model assumptions for formation and accretion. We combine these observable characteristics with survey detection thresholds to compute completeness to protoplanets, and estimate population properties, under an assumed model. Moreover, with hierarchical inference, this framework can be used to infer the model parameters themselves.

\subsection[Observational evidence for the relation between M and dotM]{Observational evidence for the relation between $M$ and $\dot M$}\label{ssec:mmdot_obs}

The relation between young stars' masses and mass accretion rates follows an empirically-established power law, $\dot M\propto M^\gamma$, $\gamma\approx 2$~\citep{Muzerolle2003, Calvet2004, Natta2006, Herczeg2008, Alcala2017, Betti2023, AlmendrosAbad2024}. The steep observed \MdtoM scaling is consistent with simple models of molecular cloud core collapse followed by viscous disk evolution \citep[e.g.][]{Dullemond2006}.

In the substellar regime, however, this relation is not as secure, and has significant scatter of 1--2~dex~\citep{Alcala2017, Betti2023, AlmendrosAbad2024}. The high scatter of the single power-law model motivated alternate ways of describing the relation at substellar masses. For example,~\citet{Muzerolle2005}, \cite{Manara2017}, and \citet{Alcala2017} find observations are best fit by a broken power law, with a steeper index $\gamma$ for \textit{isolated} brown dwarfs than stars.

Observations of accreting planetary-mass companions (PMCs)---brown dwarfs and protoplanet candidates bound to higher-mass stars---indicate higher accretion rates with weaker dependence on object mass than predicted by either single or broken power laws~\citep[e.g.][]{Sallum2015,Haffert2019, Betti2022}. This is suggestive of formation by disk fragmentation~\citep{Stamatellos2015}. To assess the effects of systematic errors on the \MdtoM scatter and the best-fit power law(s), \citet{Betti2023} compiled a database, named CASPAR, in which they re-derived mass accretion rates for 798 low-mass and substellar objects from the literature under a uniform set of stellar distance and age estimates, empirical scaling relations, and pre-main sequence evolutionary models. They find the data are best fit with a break in the slope of the \MdtoM relation at the substellar boundary and by fitting jointly for mass and age.

\subsection[Mass accretion rate to mass as a probe of formation pathways]{\MdtoM as a probe of formation pathways}\label{ssec:formpar}

A likely explanation for the fact that current data are best described by an overlapping set of \MdtoM power laws involves the data comprising multiple subpopulations defined by distinct formation conditions. An accreting object's initial environment influences the mass of its surrounding disk, which governs the relation between its mass and mass accretion rate. However, resolving separate formation channels is complicated by intrinsic accretion variability and observational uncertainty, hence substellar mass-to-mass-accretion rate power laws remain an active area of observational and theoretical research. Moreover, the observational evidence for multiple formation pathways remains entangled with systematics of accretion calculations, some of which rely on assumptions that are untested at substellar masses.

The \MdtoM power law is key for estimates of completeness because contrast is proportional to \MMd. As the nature of this relation is a subject of ongoing research, we provide a generalized framework for calculations of survey sensitivity. Sections \ref{sssec:starform}--\ref{sssec:agnostic} outline three possible approaches that cover a range of substellar formation scenarios in current literature. In this work, we compute survey sensitivity assuming each model in turn; inferring the relative importance of each model, or inferring the power-law parameters, will be the subject of future work.

\subsubsection{Empirical stellar relation}\label{sssec:starform}

A protoplanet on a stable orbit that cleared a large gap in the circumstellar disk, or one that formed as a quasi-binary companion, may have access to a limited reservoir of material for accretion set by its initial mass. For objects that formed in this manner, a reasonable proxy for the resultant \MdtoM relation is the single power law fit from~\citet{Betti2023}. The single power law relation implies formation in a more ``star-like" manner by fragmentation of a protostellar core:
\begin{equation}\label{eq:stellarformation}
\log\dot M/(M_\mathrm{J}\text{yr}^{-1}) = 2.02\log M/M_\mathrm{J}-5.00,
\end{equation} which has an associated spread of $\sigma=0.85$ dex.

This relation predicts a steep drop-off of accretion rate with mass ($\gamma\approx 2$), though with high scatter of $\approx1$ dex. Because the data underlying this fit are empirical, we emphasize that Equation~\ref{eq:stellarformation} is not derived from stellar formation physics, but instead reflects the scenario of a system with a fixed mass reservoir set early in its formation. We subsequently cite~\citet{Betti2023} as~\citetalias{Betti2023} and refer to Equation~\ref{eq:stellarformation} as the ``isolated'' or ``empirical stellar'' formation model.

\subsubsection{Theoretical planetary relation}\label{sssec:planform}

An alternate model for the formation of substellar companions is via the fragmentation of a massive gravitationally unstable circumstellar disk. Simulations of disk fragmentation suggest the resulting circumsecondary disks are more massive than independent evolution would dictate, because the disks draw matter from the natal system for longer~\citep{Stamatellos2015}. Since the accretion rate scales with disk mass, both for companions and isolated accretors~\citep{Manara2016, Somigliana2022MNRAS, Fiorellino2022}, disk fragmentation simulations predict protoplanets with higher accretion rates that are more weakly dependent on object mass than the empirical stellar relation.

Disk evolution modeling requires the disk viscosity $\alpha$, which governs the rate of angular momentum transport. As lower-viscosity disks evolve more slowly, the fragments that form circumplanetary disks pull more mass from the natal system, and their planets consequently have higher accretion rates. Indeed,~\citet{Rafikov2017} finds observational evidence for a correlation between $\alpha$ and $\dot M$. Following~\citet{Stamatellos2015}, \citet{Rafikov2017}, and \citet{Somigliana2022MNRAS}, we assume $\alpha$ does not depend on object mass and use the presumed value for T-Tauri stars, $\alpha=0.01$.\footnote{Notably, $\alpha=0.01$ agrees with the empirical fit to PMCs in~\citetalias{Betti2023}.} While viscosity impacts the vertical offset of the \MdtoM relation, it does not strongly impact the slope. As such, a different $\alpha$ would result in a constant vertical shift of the completeness maps. We further discuss the potential impact of the choice of $\alpha$ in Section~\ref{sec:conclusion}. We adopt the \MdtoM scaling with these assumptions from the simulations of~\citet[][]{Stamatellos2015}, henceforth~\citetalias{Stamatellos2015}:
\begin{equation}\label{eq:planetformation}
\log\dot M/(M_\mathrm{J}\text{yr}^{-1}) = 0.12\log M/M_\mathrm{J}-7.48,
\end{equation} which has a spread of $\sigma=0.3$ dex.

Compared to Equation~\ref{eq:stellarformation}, Equation~\ref{eq:planetformation} has a shallower slope ($\gamma=0.12$) and less variation in accretion rate ($\sigma=0.3$ dex). We subsequently refer to Equation~\ref{eq:planetformation} as the ``fragmentation'' or ``planetary'' formation model. Such a flat relation between $M$ and $\dot M$ is useful from a detection standpoint, as it implies better contrasts for low-mass objects. However, it impedes source classification and demographics: an observed $\MMd$ (computed from contrast) has more pronounced degeneracy between $M$ and $\dot M$ than if there were a stronger correlation.

\subsubsection{Formation-agnostic}\label{sssec:agnostic}

Observationally, single measurements of \Ha contrast constrain only the ratio $\MMd/R$ (see Equation~\ref{eq:lacc}). Separately constraining $M$ and $\dot M$ requires multiwavelength observations. As several GAPlanetS protoplanet candidates have only \MMd constraints, we also outline a conservative approach that avoids assuming an \MdtoM scaling. For this scenario, we consider the product $\MMd$ to be the physical parameter, which makes our completeness maps ``agnostic'' about the relation between $M$ and $\dot M$. The product $\MMd$ has been used when observing other quantities dependent on both mass and accretion, such as the flux density of dust continuum emission \citep{Shibaike2024}.

While this approach reduces one assumption, it does not enable inference about the relative frequencies of formation processes, which requires separate estimates of $M$ and $\dot M$. In anticipation of more robust protoplanet mass estimates in the future, we demonstrate computation of protoplanet survey completeness under the two \MdtoM relations outlined above, though we do not compare the GAPlanetS candidates to these results.

\subsection{Total accretion luminosity}\label{ssec:gravinfall}

In the standard model for stellar accretion, the circumstellar disk is truncated by the star's magnetic field at a radius $R_\textrm{m}$ (e.g., \citealp{Koenigl1991}). Accreting matter flows in columns along magnetic field lines from $R_\textrm{m}$ to the stellar surface at near free-fall velocity, causing an accretion shock. The resulting accretion luminosity \Lacc is given by
\begin{equation}\label{eq:lacc}
\Lacc = \frac{G\MMd}{R}\left(1-\frac{R}{R_\textrm{m}}\right).
\end{equation}

Computing \Lacc thus requires estimates for the truncation radius $R_\textrm{m}$ and object radius $R$. We use the classic estimate $R_\textrm{m} = 5R$, a value consistent with observations~\citep{Calvet1998,Hartmann2016}. While a different dependence of \Lacc on \MMd may apply for non-magnetospheric accretion processes~\citep[e.g.][]{Aoyama2021}, we note that a modification of Equation~\ref{eq:lacc} should add only an overall scale factor to our completeness estimates. We expect uncertainty in this relation to be subdominant compared to uncertainties in Equations~\ref{eq:stellarformation} and~\ref{eq:planetformation}.

We also assume a uniform value for planet radius, namely $2R_\mathrm{J}$. Alternately, planet radii can be modeled as a function of mass and mass accretion, as done in \citet{Aoyama2020} using data from \citet{Mordasini2012}. Their results agree with the population synthesis results of \citet{Emsenhuber2021}. However, \citet{Aoyama2020} find protoplanet radii vary only by a factor of two in our parameter space ($\log M/M_\mathrm{J}\in[0,2],\;\log\dot M/(M_\mathrm{J}\mathrm{yr}^{-1})<-5$), so the impact of radius uncertainty on the inferred completeness estimates is likely small. In this work, we focus on comparing models for the \MdtoM and \LHatoLacc relations. However, incorporating the \citet{Aoyama2020} fit for $R(M,\dot M)$ or comparing models for the radius relation is an extension that may better reflect planet characteristics.

\subsection[Accretion mechanisms and Halpha emission]{Accretion mechanisms and \Ha emission}\label{ssec:accpar}

Measuring \Lacc provides the most direct way to estimate $\dot M$ in principle, but it often cannot be measured in practice. Accretion emission is primarily in the near-UV continuum, with substantial line luminosity contributions in the mid-to-far UV, optical, and near-infrared. Since multiwavelength observations for low-mass and substellar objects have not yet been obtained in most cases, spectral lines such as \Ha (656 nm) are often used as secondary tracers of accretion \citep[e.g.][]{Alcala2014,Alcala2017,Natta2006,Manara2015}.
Observational work on accreting substellar objects suggests that the contribution of the accretion continuum at 656 nm is negligible compared to the line emission, as has been seen for DH~Tau~b, GQ~Lup~b, and PDS~70~b~\citep{Herczeg2008,Zhou2014,Bowler2014,Zhou2021}. However, we note \citet{Zhou2014} finds the \Ha emission from GSC~6214-210~b to be comparable to the continuum, suggesting not all companions follow this picture. Theoretical work~\citep[e.g.][]{Aoyama2021} further suggests the low contribution of the continuum. Thus, we make the simplifying assumption that observed emission in the 6~nm wide \Ha filter is pure line emission from accretion. The simultaneous differential imaging observing strategy of the GAPlanetS survey strengthens this assumption through contemporaneous continuum non-detections of accreting protoplanet candidates. Where and when this assumption is valid is discussed in detail in~\citet{Follette2023}, Section 6.2.1.

The nature of the accretion flows impacts how the total accretion luminosity is partitioned across the electromagnetic spectrum. For a given rate of accretion, how much is radiated in \Ha therefore depends on the mechanism of accretion. Since planetary accretion physics are not well constrained, we compare two models for how $\LHa$ scales with $\Lacc$: one empirically derived from observations of low-mass stars, and one theoretically derived from simulations of accretion flows onto planets.

Interpreting \Ha observations requires knowledge, or assumption, of the applicable accretion scaling relation. As identifying the scaling for any given system remains an observational challenge, we compute completeness to GAPlanets under both relations, enabling comparison of the population inference using each assumption.

\subsubsection{Empirical stellar relation}

In low-mass stars where both accretion-line emission and UV excess (as a proxy for \Lacc) can be measured simultaneously, empirical correlations have been derived for the relation between line and accretion luminosity. Here we consider the \LHatoLacc relation determined in~\citet{Alcala2017} (hereafter~\citetalias{Alcala2017}), namely:
\begin{equation}\label{eq:stellaraccretion}
\log\Lacc/L_\odot = 1.13\log\LHa/L_\odot+1.74.
\end{equation}

We note that~\citetalias{Alcala2017} recommend the use of other accretion tracers over \Ha, in part because of the high observed dispersion, thought to be due to other processes (e.g. outflows, magnetic fields) that contribute to \LHa. However, since \Ha is one of the only ground-accessible accretion tracers observable at planetary fluxes, it is a necessary, though imperfect, choice. The low-mass stars described by Equation~\ref{eq:stellaraccretion} are thought to accrete magnetospherically. During magnetospheric accretion, this line emission is believed to originate from the accretion columns, with the infalling gas producing broad emission lines. The accretion shock onto a stellar surface is almost entirely ionized, creating primarily hot continuum emission and little line emission~\citep{Hartmann2016}.

\subsubsection{Theoretical planetary relation}

In the planetary accretion case, infalling material may be too cool to emit substantially at \Ha, and the post-shock region may be cool enough to maintain bound elections. Thus line emission may originate primarily from the photospheric post-shock region, which contributes to a lower \Ha production efficiency. Additionally, planets may not have sufficiently strong magnetic fields to accrete magnetospherically, instead exhibiting ``boundary layer accretion'' (e.g.~\citealt{Fu2023}).

\citet{Aoyama2021} (hereafter~\citetalias{Aoyama2021}) quantified the expected \Ha emission of a planetary accretion shock via non-equilibrium radiation-hydrodynamic simulations combined with a simple estimate for the accretion geometry and found: \begin{equation}\label{eq:planetaccretion}
\log\Lacc/L_\odot = 0.95\log\LHa/L_\odot+1.61,
\end{equation}
which, compared to Equation~\ref{eq:stellaraccretion}, predicts \textit{less} \Ha emission for a given \Lacc.

The fit of Equation~\ref{eq:planetaccretion} was obtained in the limit of azimuthally-symmetric accretion onto a planet, assuming a shock only at the planetary surface. In a more realistic accretion geometry (e.g.,~\citealp{Tanigawa2012}), the circumplanetary disk surface shock will also add to the line flux~\citep{Aoyama2018}, but its contribution is likely subdominant compared to the photospheric shock~\citep{Marleau2023,Takasao2021}.

While the scaling relations Equations ~\ref{eq:stellaraccretion} and~\ref{eq:planetaccretion} nominally delineate ``stellar'' and ``planetary'' accretion processes, either mechanism may apply for a given substellar object. Indeed, observations of PDS~70~b imply a high fraction of accretion emission at \Ha, with such efficient production suggesting a more ``stellar'' accretion process~\citep{Zhou2021}. Upcoming studies of the PDS~70 system hope to further unveil the process at play~\citep{Aoyama2022hst}. Similarly, measurements of the number density of the accretion flow onto Delorme~1~(AB)b imply a small line-emitting area, suggesting the accretion flow is a column~\citep{Betti2022,Ringqvist2023}.

\subsection{Modeling planetary orbits}\label{ssec:orbits}

For consistency with previous work (e.g.~\citealt{Nielsen2019}), we compute detectability as a function of semimajor axis marginalized over other orbital parameters. To model companions' orbits, we use the typical assumption of isotropy in viewing angle and most orbital characteristics: we model inclination angle $i$ as uniform over $\cos(i)\in[-1,1]$; argument of periastron $\omega$ as uniform in $\omega\in[0,2\pi]$; and epoch of periastron passage $\tau$, expressed as fraction of orbital period past a reference time, as uniform over $\tau\in[0,1]$.  The position angle of nodes, which sets an orbit's azimuthal orientation, is not simulated, as GAPlanetS contrast curves are azimuthally averaged.

The eccentricity distribution of exoplanets is less well constrained. Planet population studies have employed a range of possibilities, and recent studies have put new constraints on the distribution. \citet{Nielsen2019} uses $P(e)\propto 2.1-2.2e,\,0\leq e\leq0.95$, derived from the radial velocity survey of~\citet{Butler2006}. \citet{Bowler2015} compares results using circular orbits ($e=0$) to those using a nearly-identical distribution to \citet{Nielsen2019}, $P(e)\propto 1-e$, which is based on a combination of radial velocity and M dwarf binary observations~\citep{Duchene2013,Kipping2013}. More recent work by \citet{Bowler2020} fit the eccentricities of imaged giant planets and brown dwarfs to beta distributions and found a dependence on companion mass and orbital period. They interpret these trends as evidence for different formation channels, suggesting the proper eccentricity model may depend on the formation condition. The population study of \citet{Vigan2021} uses the best-fit parameters for the entire sample of \citet{Bowler2020} ($\alpha=0.95$, $\beta = 1.30$). \citet{DoO2023} fits a sub-sample of objects from \citet{Bowler2020}, finding Beta parameters $\alpha=1.09$, $\beta = 1.42$---a nearly uniform distribution. \citet{Nagpal2023} devises a hierarchical Bayesian method to measure the population-level eccentricity distribution and finds $\alpha=0.7_{-0.3}^{+0.4}$, $\beta = 2.3_{-0.7}^{+0.8}$, which is broadly consistent with $P(e) \propto 2.1-2.2e$ to within uncertainty. \citet{Wahhaj2024} finds $e\approx0.5$ for PDS~70~b and c, suggesting high eccentricities are possible for protoplanets.

To assess the importance of the assumed eccentricity distribution, we compare two possible distributions. As a conservative possibility, we assume all orbits have circularized; i.e., $P(e) = \delta_{0,e}$. As a second possibility, we follow \citet{Nielsen2019} and use $P(e)\propto 2.1-2.2e,\,0\leq e\leq0.95$.

With the models for \MdtoM and \LHatoLacc in  Sections \ref{ssec:formpar}--\ref{ssec:accpar}, we perform the schematic mapping of Equation~\ref{eq:mapping} and compute the luminosity of a protoplanet. Using the orbital models here, we compute the projected separation. This completes the requirements for translating physical to observable parameters. In the remainder of this article, we apply this framework to real observations from GAPlanetS~\citep{Follette2023} to compute the survey completeness.

\section{Data} \label{sec:data}

The GAPlanetS collected \Ha images of 14 transitional disk systems with the Magellan Adaptive Optics~\citep{Close2013,Morzinski2014,Morzinski2016} system. THE GAPlanetS sample selection, data collection, reduction, and analysis pipelines are detailed in~\citet{Follette2023} and~\citet{AdamsRedai2023}.

Since our aim is planet demographics, we must define the stellar population on which we are performing inference. GAPlanetS is a highly targeted survey: it selected stars with transitional disks hosting large ($>0\farcs1$), cleared central cavities visible from the Magellan telescope ($\delta<25°$) and of sufficient brightness for observation with natural guide star adaptive optics (rmag$<$12). While GAPlanetS does not explicitly exclude known binary systems, these criteria do rule out several well-studied circumbinary disks, such as GG~Tau. Our inference about protoplanet populations is thus limited to gapped transitional disk host stars. While this is a sampling bias, it is also a consequence of the fact that we expect forming planets to be most prevalent inside disk gaps and cavities.

GAPlanetS reported six detections of accreting companions and candidates in five systems. Two (HD~142527~B, HD~100453~B) are low-mass stars. Although not protoplanets, their formation environments may be analogous to substellar companions, which motivates studying their formation and accretion properties~\citep{Balmer2022}. GAPlanets recovered the confirmed protoplanet PDS~70~c and candidates LkCa~15~b and CS~Cha~c. They did not robustly recover PDS~70~b, so we exclude it from this analysis.

It computed the detection threshold as a function of separation, or contrast curve, for each epoch of observation for each star. Contrast curves are determined via the injection and recovery of false planets at fixed contrasts. After injection, the stellar point-spread function (PSF) is modeled and subtracted via {\tt pyKLIP}~\citep{Wang2015} using the optimization process described in \citet{AdamsRedai2023} and \citet{Follette2023}. In these post-processed images, the detection threshold is set using the remaining 5$\sigma$ image noise corrected for small sample statistics at low separations~\citep{Mawet2014} and multiplied by the algorithm throughput (the ratio between injected and post-{\tt pyKLIP} planet flux). As a caveat, we note that the method of \citet{Mawet2014} was generalized to non-Gaussian noise in \citet{Bonse2023}. They find an incorrect assumption about the noise distribution can impact the derived contrast curve by about 1 magnitude at close separations, which may affect detections within two full widths at half max of the star. This does not affect GAPlanetS detections, but increases uncertainty for completeness estimates at low separations.

The relevant stellar parameters for computing completeness for each survey star are given in Table~\ref{tab:surveystars}. We compare the simulated companions to the optimized contrast curves for each survey star. This gives the detection probability for each star as a function of companion parameters, conditioned on the astrophysical models.

\begin{table*}
\begin{tabular}{cccccc}
\hline \hline
Star & $M$ & $d$\textsuperscript{1} & $m_{r',\star}$\textsuperscript{1,2} & $A_{r',\star}$\textsuperscript{1,2,3} & $S$ \\
& [$M_\odot$] & [pc] & [mag] & [mag] & \\\hline
HD~100546 & 2.2\textsuperscript{4} & 108.1 & 6.8 & 0.2 & 1.59, 1.43 \\
HD~141569 & 2.39\textsuperscript{5} & 111.6 & 7.2 & 0.2 & 0.94, 0.96, 0.95, 0.92, 0.92 \\
HD~100453 & 1.7\textsuperscript{6,7} & 103.8 & 7.8 & 0.2 & 1.05, 1.26, 1.04, 1.10 \\
HD~142527 & 2.0\textsuperscript{8} & 159.3 & 8.2 & 0.8 & 1.13, 0.88, 1.12, 1.14, 1.22, 1.28, 1.13, 1.14 \\
HD~169142 & 1.85\textsuperscript{9} & 114.9 & 8.2 & 0.0 & 1.06, 0.98, 0.99, 1.13 \\
SAO~206462 & 1.7\textsuperscript{10} & 135.0 & 8.6 & 0.1 & 1.22, 1.22 \\
V1247~Ori & 1.86\textsuperscript{11} & 401.3 & 9.9 & 0.3 & 1.13, 1.12 \\
PDS~66 & 1.4\textsuperscript{12} & 97.9 & 10.0 & 0.7 & 1.91 \\
V4046~Sgr & 0.9, 0.85\textsuperscript{13} & 71.5 & 10.0 & 0.0 & 1.80, 1.79 \\
TW~Hya & 0.6\textsuperscript{14} & 60.1 & 10.5 & 0.5 & 8.79, 7.17 \\
CS~Cha & 1.32\textsuperscript{15} & 168.8 & 11.1 & 1.0 & 2.26 \\
UX~Tau~A & 1.20\textsuperscript{16} & 142.2 & 11.3 & 0.5 & 1.42 \\
LkCa~15 & 1.25\textsuperscript{17} & 157.2 & 11.6 & 0.5 & 1.81, 1.58 \\
PDS~70 & 0.82\textsuperscript{18} & 112.4 & 11.7 & 0.0 & 1.29, 1.36, 1.32 \\
\hline\hline
\end{tabular}
\caption{Selected parameters of GAPlanetS survey stars. From left to right: stellar mass in solar masses, distance in parsecs, apparent $r'$-band magnitude, $r'$-band extinction, and \Ha-to-continuum scale factor (see Section~\ref{ssec:detec}) for each observational epoch, ordered by epoch. Additional parameters and derivations are given in Tables~1, 2, and~5 and Section~6 of~\citet{Follette2023}. References: \textsuperscript{1}\citet{Gaia2023}, \textsuperscript{2}\citet{Alam2015}, \textsuperscript{3}\citet{Pecaut2013}, \textsuperscript{4}\citet{Casassus2022}, \textsuperscript{5}\citet{White2016}, \textsuperscript{6}\citet{Dominik2003}, \textsuperscript{7}\citet{Collins2009}, \textsuperscript{8}\citet{Mendigutia2014}, \textsuperscript{9}\citet{Gratton2019}, \textsuperscript{10}\citet{Muller2011}, \textsuperscript{11}\citet{Kraus2013}, \textsuperscript{12}\citet{Avenhaus2018}, \textsuperscript{13}\citet{Stempels2004}, \textsuperscript{14}\citet{Sokal2018}, \textsuperscript{15}\citet{Manara2014}, \textsuperscript{16}\citet{Kraus2009}, \textsuperscript{17}\citet{Donati2019}, \textsuperscript{18}\citet{Riaud2006}.}
\label{tab:surveystars}
\end{table*}

\section{Method for computing completeness}\label{sec:methods}

We determine GAPlanetS' completeness to accreting companions following a Monte Carlo procedure. For each target star, we compute the completeness on a 60x60 grid. We use a log-uniform grid in semimajor axis ranging from 1 to 500 au. For the $y$-axis, we use either the log (base 10) of mass in the range $\log M/M_\mathrm{J}\in[0,2.5]$, or the log of the product $\MMd$, $\log \MMd/(\MMdunits)\in[-10,-2]$. These ranges span planets, brown dwarfs, and the lowest-mass stars. At each set of parameters $(a, M)$ or $(a, \MMd)$, we simulate $10^4$ companions. We determine the detectable fraction by estimating the observable parameter distribution $P(\mathrm{separation},\LHa)$ for the given physical parameters. We do so by simulating companion orbits and \LHa, the latter of which depends on the scaling law for each step in Equation~\ref{eq:mapping}. The completeness maps suppose equal probability in each bin, which corresponds to log-uniform priors on semimajor axis and mass.

The steps in our simulations are diagrammed as a flowchart in Figure~\ref{fig:flowchart}. We first select models for \MdtoM, \LHatoLacc, and orbits; we discuss our choices in detail in Sections \ref{ssec:formpar}-\ref{ssec:orbits}. For each survey star, we then simulate $10^4$ companions. Although protoplanet detections in accretion tracing lines place minimal direct constraints on planet mass, we can estimate mass limits from accretion rate limits if we assume a relation between $M$ and $\dot M$. We consider one ``star-like" \citepalias[Equation \ref{eq:stellarformation},][]{Betti2023} and one ``planet-like" model \citepalias[Equation \ref{eq:planetformation},][]{Stamatellos2015}, shown in purple and green, respectively, in panel (A) of Figure~\ref{fig:flowchart}. We also conservatively directly simulate $\MMd$, avoiding assuming a relation. In all cases, we convert from \MMd to \Lacc with Equation~\ref{eq:lacc}. Next, we compute \LHa assuming either a more ``star-like" \citepalias[Equation~\ref{eq:stellaraccretion},][]{Alcala2017} or ``planet-like" \citepalias[Equation~\ref{eq:planetaccretion},][]{Aoyama2021} accretion mechanism, again shown in purple and green in panel (B). We randomly assign orbital parameters to compute the distribution of projected separation (panel (C)). To compute the contrast of each simulated companion, we use the stellar \Ha luminosities from the GAPlanetS survey. Finally, we label each simulated companion as ``detectable'' or ``undetectable'' in GAPlanetS data, by comparing its separation and contrast to the target star's contrast curve (panel (D)). We further describe the contrast computation below. We repeat across our grid in separation and \MMd for each survey star and each combination of models.

\subsection{Detectability}\label{ssec:detec}

Following \citet{Follette2023}, we compute each simulated companion's contrast via \begin{equation}\label{eq:contrast}
    \log C = \log\frac{\LHa}{4\pi d^2 z\Delta\lambda} + \frac{m_{r',\star}-A_{r',*}}{2.5}-\log S,
\end{equation}
where $z$ is the instrumental zero point; $\Delta\lambda$ the effective filter width; $d$ the distance to the star; $m_{r',\star}$ and $A_{r',\star}$ the star's apparent magnitude and extinction, respectively, in $r'$ band, which is a good proxy for the \Ha continuum values; and $S$ the star's scale factor, the \Ha/continuum ratio for the host star, which is used in image reduction to aid the removal of stellar residuals and scattered light. For the Magellan Clay telescope, used by GAPlanetS, $z = 1.733\times10^{-5}\;\text{erg cm}^{-2}\text{ s}^{-1}\;\upmu\text{m}^{-1}$ and $\Delta\lambda = 6.3$~nm~\citep{Males2013}. The values of each parameter for each star is provided in Table~\ref{tab:surveystars}. Details of the derivations, and additional parameters, are in Tables~1, 2 and~5 and Section~6 of~\citet{Follette2023}.

We compare each set of simulated companions to each survey star's contrast curve(s). Many of the survey stars were observed in multiple epochs. Since contrast curves vary between observation nights, a companion may be revealed or drop under the threshold. If a simulated companion is above at least one contrast curve, it is labeled ``detectable''. We do not evolve the companions' orbits between epochs. Since the distribution of projected separation will not change with orbital evolution, it will not affect the aggregate detectable fraction. Further, accretion rates are known to vary on short and long timescales. Since the stochastic accretion rate variability is unknown, we cannot deterministically predict how an object's assigned \MMd may change between epochs; however, the overall distribution of \MMd should again be unchanged. Two representative examples of simulated companions at a single $(a,M)$ are provided in panel~(D) of Figure~\ref{fig:flowchart}. The objects above the contrast curve and between the inner and outer working angles are detectable; those below the curve or outside the separation bounds are undetectable. Each star's completeness map stores the fraction of companions detectable at each $(a,M)$. Panel~(E) shows how the simulations at each $(a,M)$ are ``collapsed'' into a single grid point in the overall map. The sum across stars gives the total survey completeness to companions. This figure of merit for GAPlanetS is shown and discussed in Section~\ref{sec:completeness}. 

\section{GAPlanetS completeness to accreting companions}\label{sec:completeness}

\begin{figure*}[!htb]
    \centering
    \includegraphics[width=0.98\textwidth]{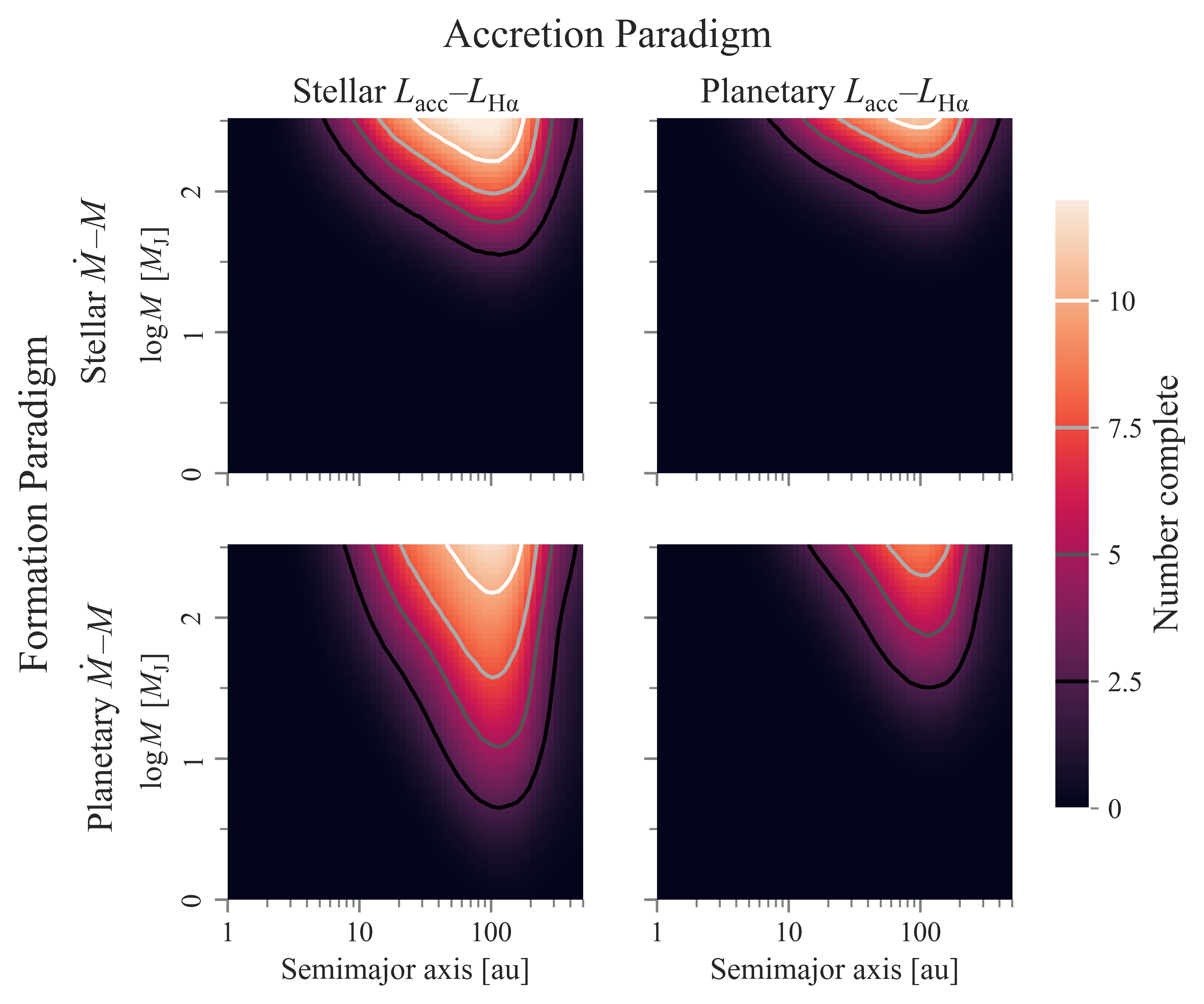}
    \caption{Completeness to accreting companions for the 14 stars surveyed by GAPlanetS as a function of semimajor axis (astronomical units) and mass $(M_\mathrm{J})$ for the four combinations of models. The color scale shows the depth of search or the number of stars to which the survey is complete. The rows dictate the \MdtoM relation, with the top being the stellar~\citepalias{Betti2023} model and bottom being the planetary~\citepalias{Stamatellos2015} model. The columns define the \LHatoLacc scaling, with the left being the stellar~\citepalias{Alcala2017} and right being the planetary~\citepalias{Aoyama2021} scalings. The contours show the completeness to 2.5, 5, 7.5, and 10 (out of 14) stars in the sample.}
    \label{fig:completeness}
\end{figure*}

We apply the method detailed in Section~\ref{sec:methods} to the full GAPlanetS dataset. Figure~\ref{fig:completeness} provides the survey-averaged completeness to accreting companions as a function of semimajor axis and mass for the four combinations of accretion and formation models we consider. Across models, the smallest mass to which the survey is at least 25\% complete is $\sim 7\;M_\mathrm{J}$. This deepest sensitivity occurs when assuming the planetary formation and stellar accretion models (lower left panel of Figure~\ref{fig:completeness}) at semimajor axes of $\sim80$--110~au.

The trends in detectability in the four panels of Figure~\ref{fig:completeness} follow those anticipated by the differences in the models. Since the planetary \MdtoM relation predicts higher $\dot M$ values at planetary masses than the stellar relation, the survey reaches its deepest sensitivity assuming planet-like formation. Similarly, since the stellar $\LHatoLacc$ accretion scaling predicts more emission in \Ha for a given $\Lacc$ than the planetary scaling, the left column has deeper sensitivity than the right. However, the impact of the accretion scaling is less significant when using the stellar formation relation. This is due to the larger variance of accretion rates at a given mass, which tends to smooth out the effect of the accretion scaling (see panel~(A) of Figure~\ref{fig:flowchart}, where the higher $\sigma$ of the \citetalias{Betti2023} relation than the \citetalias{Stamatellos2015} relation is clear).

While intuition suggests cross-combinations of one ``planetary" and one ``stellar" model would be less likely, current data suggest any of the four pairs of models shown here are possible. For instance, observations of Delorme~1~(AB)b suggest the presence of concentrated accretion columns (i.e., the stellar model), but a high disk mass indicative of formation by fragmentation~\citep{Ringqvist2023}. Importantly, we are most complete to objects that underwent ``stellar" accretion and ``planetary" formation, so such objects are more easily detected than objects formed by other combinations of models, even though they may be less astrophysically abundant. As another example, the PMC TWA~27~B has a mass ratio $q\approx0.2$ with its host star, implying the system formed more like a stellar binary and its accretion properties are consistent with both the magnetospheric and planetary shock models~\citep{Marleau2024,Aoyama2024}. PDS~70~b is fit well by the magnetospheric accretion model~\citep{Thanathibodee2019}, though it is certainly of planetary mass, and its accretion rate varies on $\lesssim1$ year timescales~\citep{Close2025}. These empirical examples demonstrate that although current models cannot fully explain how such cross-combinations occur, there is growing evidence that a strict binary of ``planet" versus ``stellar" formation and evolution is insufficient.

Identifying the formation and accretion channels of a given object typically requires multi-epoch, multiwavelength observations, which provide better measurements of $M$ and $\dot M$. While we limit this analysis to the \Ha results of GAPlanetS, future work will combine studies across wavelengths for stronger constraints.

\subsection{Impact of eccentricity}\label{ssec:ecc}

\begin{figure}[!htb]
    \centering
\includegraphics[width=0.48\textwidth]{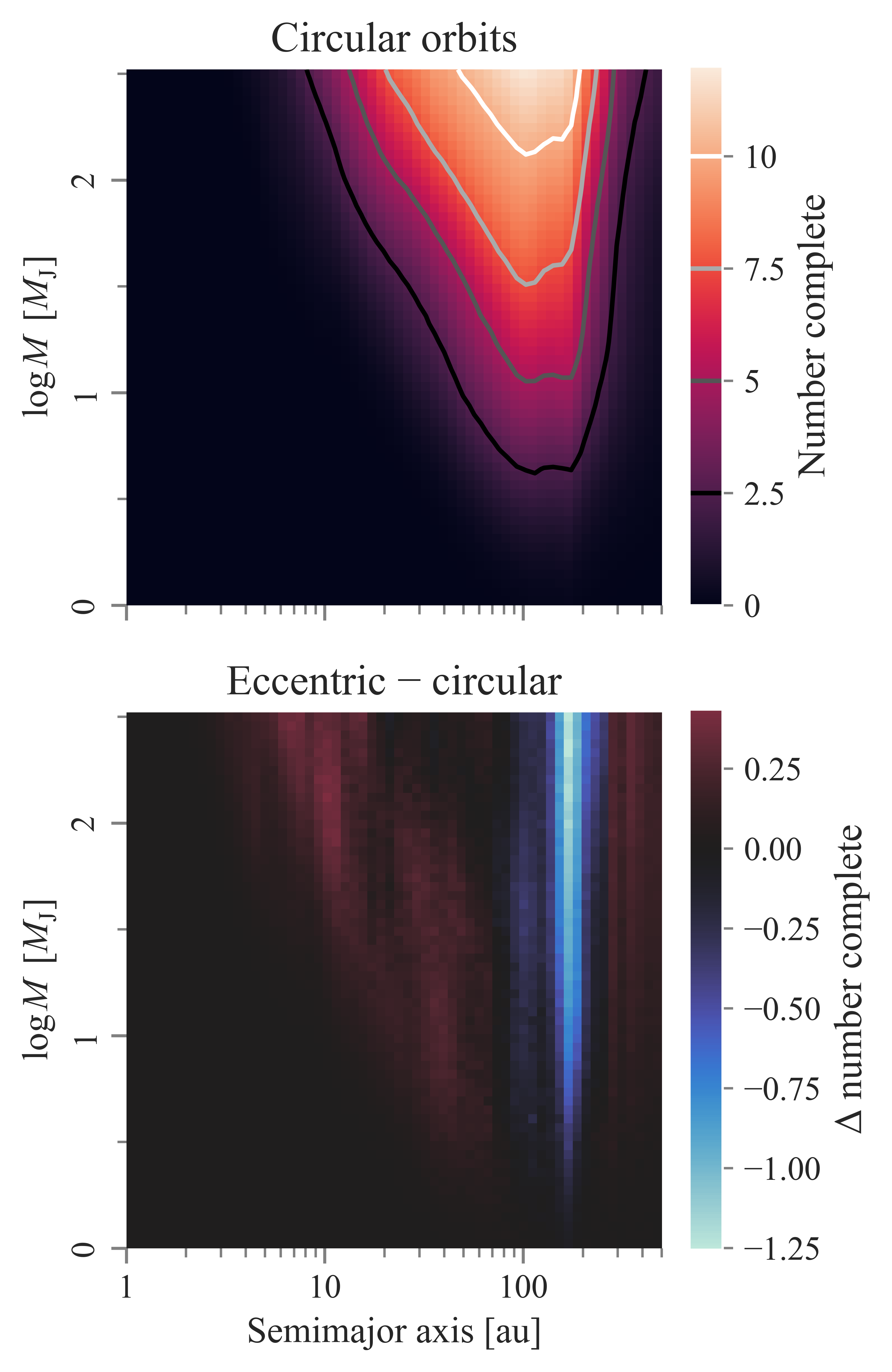}
    \caption{Top: GAPlanetS completeness to companions assuming circular orbits, $P(e)=\delta_{0,e}$, using the planetary \MdtoM and stellar \LHatoLacc scalings. The color scale and contours are as in Figure \ref{fig:completeness}. \textit{Bottom:} difference in survey sensitivity when assuming orbital eccentricities follow $P(e)\propto 2.1-2.2e$ (lower left panel of Figure~\ref{fig:completeness}) versus assuming circular orbits, $P(e)=\delta_{0,e}$. The colorbar shows the difference in the number of stars to which the survey is complete (out of 14); positive values indicate eccentric companions are easier to detect.}
    \label{fig:eccdistr}
\end{figure}

To assess the effect of the assumed eccentricity distribution on sensitivity, we compare the completeness maps under fixed formation and accretion scaling relations but with eccentricities from $P(e)\propto 2.1-2.2e$ versus circular orbits, $P(e)=\delta_{0,e}$. Figure~\ref{fig:eccdistr} compares the results under the planetary formation and stellar accretion relation. The top panel shows the map assuming circular orbits, while the lower panel plots the difference between the map allowing for eccentric orbits (the lower left panel of Figure~\ref{fig:completeness}) and the purely circular map. While the maps are qualitatively similar in shape, the map assuming circular orbits has more structure, because not including eccentricity limits the spread of the projected separation.

Eccentricity has the largest effect on detectability at wide semimajor axes, where circular orbit companions are easier to detect. This is because eccentric companions are more likely to be projected beyond the outer working angle, decreasing detectability. At low semimajor axes, it is marginally easier to detect eccentric companions. Because the detectability threshold is highest at close separations, low-semimajor-axis planets are easier to detect when are eccentric and are, on average, farther from their host than their semimajor axis.

The integrated completeness is nearly identical between the eccentric and non-eccentric maps, and changes at most by 1.25 stars out of 14 (roughly $9\%$) at wide separations. Since planet demographics show giant planets are rare at wide separations (see Section~\ref{ssec:separation} and references therein), the behavior toward small semimajor axes, where ecccentricity has a smaller effect, is more influential for occurrence rate estimates. Moreover, the outer working angle is an artifact of the chosen telescope. Many of these systems have been imaged with classical high-contrast imaging surveys with larger limits~\citep[e.g., ][]{Nielsen2019,Vigan2021}, so any such high-eccentricity, wide-separation planets likely would have been detected through other surveys. The assumption about planet eccentricity is a subdominant effect in completeness and occurrence-rate estimates compared to the formation and accretion models.

\section{Constraints on protoplanet occurrence rate}\label{sec:constraints}

\begin{figure*}[!htb]
    \centering
    \includegraphics[width=0.98\textwidth]{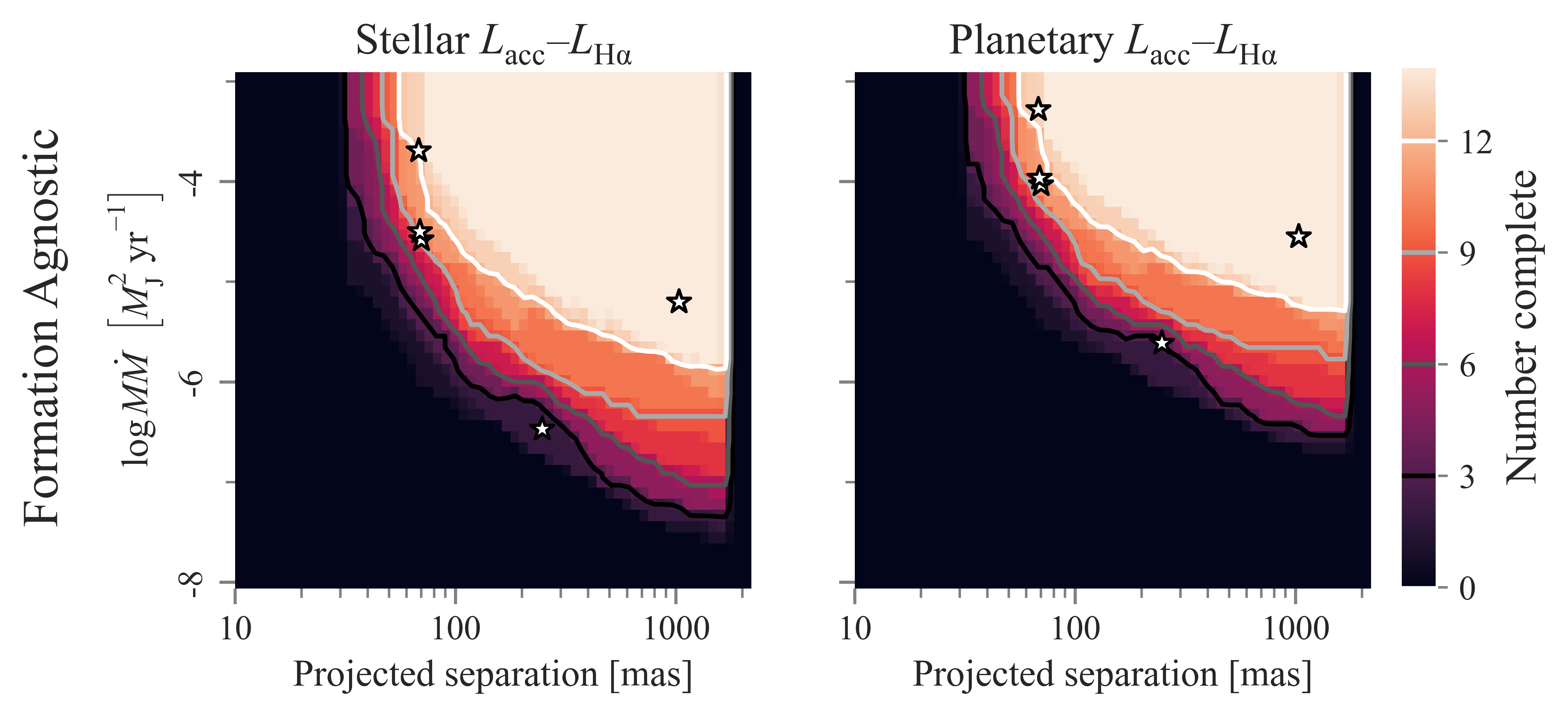}
    \caption{Survey-averaged completeness to accreting companions as a function of projected separation (mas) and \MMd (\MMdunits). The five GAPlanetS detections are overplotted as white stars. From upper left to lower right, these are CS~Cha~c, LkCa~15~b, HD~142527~B, PDS~70~c, and HD~100453~B.}
    \label{fig:completeness_withdets}
\end{figure*}

To extend our completeness framework to constraints on protoplanet populations, we adopt the most conservative approach and do not rely on an assumption of the \MdtoM relation, instead simulating \MMd as the physical parameter as covered in Section~\ref{sssec:agnostic}. Since we have precise constraints on separation but not semimajor axis, we also marginalize over semimajor axis to get the completeness as a function of separation. When projecting into separation space, we assume a log-uniform prior in semimajor axis, $dN/da\propto a^{-1}$, as in~\citet{Nielsen2019} and~\citet{Vigan2021}. From the \Ha contrasts of the five GAPlanetS candidate protoplanet detections, we compute $\MMd$ under both accretion scalings. For HD~142527~B, which was observed in six epochs, we use the average separation and inferred $\MMd$. We plot the detections over the completeness maps in Figure~\ref{fig:completeness_withdets}. Although we simulated $\log\MMd/(\MMdunits)\in[-10,-2]$, we truncate the plot at $-8$, below which the survey has zero sensitivity under either accretion relation. We find the survey is roughly one order of magnitude in \MMd more sensitive to companions that accrete magnetospherically. Planet-like accretors are detectable down to $\log\MMd/(\MMdunits)=-6$ at the widest separations. Although the intention of simulating \MMd is to avoid assuming a scaling between mass and mass accretion rate, we can estimate a limit in mass from the limit in \MMd using a chosen \MdtoM relation. We find the smallest mass to which the survey may be sensitive is $\sim 10\ M_\mathrm{J}$, consistent with the results in Figure~\ref{fig:completeness}.

We then compute posteriors on the proportion of transitional disk stars with accreting companions. We do so for different subsets of the detected companions, to analyze whether the estimate companion rate depends on either the companion or host star properties. In turn, we consider accreting companions of any mass and location, accreting companions within the transitional disk cavity, and confirmed protoplanets.

Two key, interconnected analysis choices are the range of parameter space in which we infer the rate and the priors we assume on the underlying separation and \MMd distributions. If we include regions with near-zero sensitivity, our occurrence rates involve extrapolation from the areas to which we are complete, according to the separation and mass priors we assume. Although we expect close, low-mass, weakly accreting companions to be the most common, GAPlanetS is least complete to such objects. The steep priors on separation and \MMd amplify the importance of this region of incompleteness, where the survey is entirely uninformative.

Because the chosen parameter range and priors profoundly affect the resultant rates, we analyze the data with several options for each. We present fiducial ranges and priors below and include summary statistics for alternate choices in Table~\ref{tab:rate_summaries}. As anticipated, for the same prior, restricting the space of inference to regions of higher completeness decreases the inferred rates. For the same parameter range, steeper priors tend to increase the rates.

We first consider accreting objects at separations 0$\farcs$03--2" and $-7\leq\log\MMd/(\MMdunits)\leq-3$. Although we simulated a broader range of \MMd, we exclude the extremal regions for two distinct reasons. First, we cut below 0$\farcs$03 and $\log\MMd/(\MMdunits)=-7$, as the survey has near zero sensitivity in those regions, especially for the planetary accretion case. Second, we cut the highest \MMd, as such objects would not correspond to PMCs even under the most optimistic assumptions about the \MdtoM scaling (see panel (A) in Figure~\ref{fig:flowchart}). Indeed, Figure~\ref{fig:completeness_withdets} shows a lack of detections above $\log\MMd/(\MMdunits)=-3$, suggesting companions in that regime are rare.

For our primary analyses, we set log-uniform priors in projected separation and \MMd:
\begin{align}
    P(\mathrm{sep})&\propto\mathrm{sep}^{-1};\\
    P(\MMd)&\propto\MMd^{-1}.
\end{align}
The total completeness to accreting objects, or overall search depth, is found by integrating the completeness map over the parameter space, weighted by the prior for each bin. Under these priors, GAPlanetS is complete to \complst/14 stars under the stellar accretion relation and \complpl/14 stars under the planetary relation. The log-uniform priors reflect our assumption that close, less massive, weakly accreting planets are most common. We note that the fits for the power-law indices of planet mass and separation spectra in \citet{Nielsen2019} suggest steeper drop-offs. As the survey is most complete to wide, heavily accreting companions, analyses using steeper power-law indices imply lower survey completeness, and, consequently, higher companion rates. We provide posterior occurrence rates using alternate prior choices in Table~\ref{tab:rate_summaries}.

We compute posteriors on accreting object occurrence rate using the Bayesian method described in Appendix~\ref{app:stats}. In the analyses presented below, we exclude the candidate companions LkCa~15~b and CS~Cha~c, to be more conservative about the accreting companion rate.

\begin{table*}[tp]
\centering
\begin{tabular}{c|c|c|cc|cc|cc}
\hline\hline
Integration range & \MMd prior & Separation &\multicolumn{2}{c|}{Accreting companion} & \multicolumn{2}{c|}{In disk gap} & \multicolumn{2}{c}{Protoplanet} \\
$\left(\log\MMd\times \mathrm{sep}\right)$ & index & prior index & Stellar & Planetary & Stellar & Planetary & Stellar & Planetary \\ \hline
$[-7, -3] \times [30, 2000]$ & $-1$ & $-1$ & $0.43^{+0.28}_{-0.19}$ & $0.57^{+0.36}_{-0.26}$ & $0.29^{+0.24}_{-0.15}$ & $0.40^{+0.32}_{-0.21}$ & $0.16^{+0.19}_{-0.10}$ & $0.22^{+0.26}_{-0.14}$ \\
$[-8, -3] \times [30, 2000]$ & $-1$ & $-1$ & $0.52^{+0.33}_{-0.23}$ & $0.68^{+0.39}_{-0.31}$ & $0.36^{+0.29}_{-0.19}$ & $0.48^{+0.38}_{-0.25}$ & $0.19^{+0.23}_{-0.12}$ & $0.27^{+0.31}_{-0.17}$ \\
$\:\: [-6, -3] \times [30, 2000]$\textsuperscript{a} & $-1$ & $-1$ & $0.24^{+0.20}_{-0.13}$ & $0.45^{+0.29}_{-0.20}$ & $0.13^{+0.15}_{-0.08}$ & $0.31^{+0.25}_{-0.16}$ & $0.03^{+0.09}_{-0.02}$ & $0.17^{+0.20}_{-0.11}$ \\
$[-7, -3] \times [30, 2000]$ & $-1$ & $-2$ & $0.74^{+0.40}_{-0.33}$ & $0.92^{+0.37}_{-0.38}$ & $0.53^{+0.40}_{-0.28}$ & $0.71^{+0.44}_{-0.36}$ & $0.30^{+0.34}_{-0.19}$ & $0.42^{+0.45}_{-0.27}$ \\
$[-7, -3] \times [60, 2000]$ & $-1$ & $-2$ & $0.45^{+0.30}_{-0.21}$ & $0.63^{+0.39}_{-0.28}$ & $0.31^{+0.26}_{-0.16}$ & $0.44^{+0.36}_{-0.23}$ & $0.17^{+0.20}_{-0.11}$ & $0.24^{+0.28}_{-0.16}$ \\
$[-7, -3] \times [60, 2000]$ & $-1.5$ & $-1.5$ & $0.85^{+0.39}_{-0.36}$ & $1.12^{+0.27}_{-0.39}$ & $0.63^{+0.44}_{-0.33}$ & $0.96^{+0.37}_{-0.43}$ & $0.36^{+0.40}_{-0.23}$ & $0.67^{+0.50}_{-0.41}$  \\ \hline\hline
\end{tabular}
\caption{Accreting companion occurrence rates from GAPlanetS for different parameter space integration ranges and priors on the underlying distributions of companions. The parameter space range is given in $\log\MMd/(\MMdunits)\times\mathrm{separation \ [mas]}$. The distributions of \MMd and separation are modeled as power laws, with the spectral index as listed. The subsequent columns provide the posteriors on the rates, given as medians and 68\% CIs, for three subtypes of companions (accreting companions, those in transitional disk cavities, and protoplanets) and two accretion scaling relations (Section~\ref{ssec:accpar}). \textsuperscript{a} We note that under the stellar accretion scaling relation, PDS~70~c has $\log\MMd/(\MMdunits)<-6$. So the single protoplanet falls outside the chosen parameter space and the rates under the stellar relation exclude this detection.}\label{tab:rate_summaries}
\end{table*}

\begin{figure*}[!htb]
    \centering
    \includegraphics[width=0.95\textwidth]{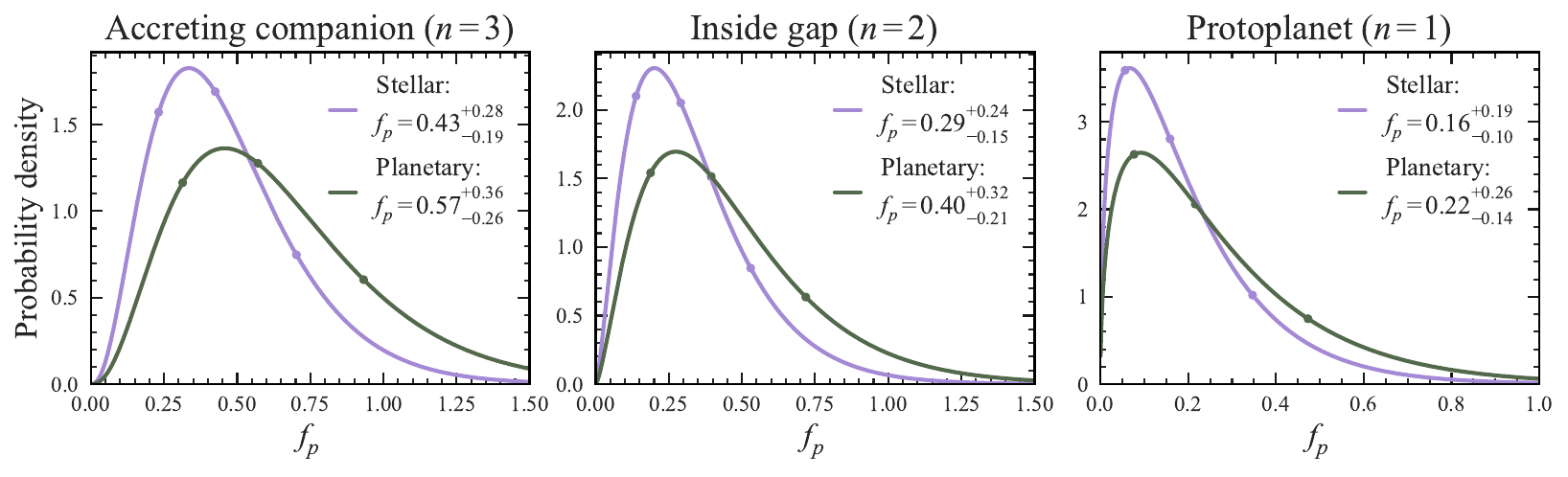}
    \caption{Posteriors on the accreting object occurrence rate around stars hosting transitional disks, $f_p$, for different subsets of the GAPlanetS detections. In all, we compare the stellar (purple) and planetary (green) accretion scaling relations. The legends give the median and 16th and 84th percentiles, also marked by filled circles. Left: posteriors on $f_p$ using the three confident detections. Center: posteriors on the rate of accreting companions \textit{inside} disk gaps, excluding HD~100453~B, which lies outside the gap. Right: posteriors on the accreting protoplanet rate, excluding the low-mass stars HD~100453~B and HD~142527~B.}
    \label{fig:posteriors_conf}
\end{figure*}

\subsection{Posteriors on occurrence rates}

The left panel of Figure~\ref{fig:posteriors_conf} shows the posterior distribution on the fraction of transitional disk systems hosting accreting objects in the parameter range 0$\farcs$03--2" and $-7\leq\log\MMd/(\MMdunits)\leq-3$, assuming the objects accrete by the stellar (purple) or planetary (green) accretion scaling relation. We label this quantity $f_p$ and note that it can be above unity (above 100\%) if the average star hosts more than one accreting companion. Since planet-like accretors should be harder to detect (see Section~\ref{ssec:accpar} and panel~(C) of Figure~\ref{fig:flowchart}), if we assume the GAPlanetS detections accrete via the planetary model, we infer a higher underlying rate of accreting objects. Indeed, the median $f_p$ is \mediancompratestell under the stellar model and \mediancomprateplan using the planetary one. The maximum likelihood values are \MLcompstel and \MLcompplan, respectively. However, the inferred companion rates are consistent to well within uncertainty. The 1$\sigma$ credible intervals (CIs) of \CIcompstel (stellar model) and \CIcompplan (planetary model) are consistent and broad, owing to small sample statistics.

The distributions in the leftmost panel of Figure~\ref{fig:posteriors_conf} include two companions known to be low-mass stars: HD~142527~B and HD~100453~B. While HD~142527~B is inside the transitional disk cavity, HD~100453~B is an external perturber that is outside the main disk gap. Although HD~142527~B is a star, its formation and accretion properties may be more similar to those of PDS~70~c, and planets and brown dwarfs that formed inside disk gaps, than to HD~100453~B and other stars. Nonembedded accreting companions, like HD~100453~B and Delorme~1~(AB)b, may have distinct population properties. Since the properties of accreting companions in and outside transitional disk gaps may differ, we also compute the rate solely for the objects inside the gaps, shown in the center panel of Figure~\ref{fig:posteriors_conf}. Assuming the stellar (planetary) accretion model, the median and 1$\sigma$ CI on $f_p$ is \ingapratestelluncert (\ingaprateplanuncert), with maximum likelihood value $f_p=\MLingapstel \ (\MLingapplan)$.

We can also consider only the bona fide accreting protoplanet PDS~70~c. In the rightmost panel of Figure~\ref{fig:posteriors_conf}, we show posteriors on the rate of accreting protoplanets using this single detection. Supposing PDS~70~c accretes via the magnetospheric (planetary) model, the median and 1$\sigma$ CI on $f_p$ is \ppratestelluncert (\ingaprateplanuncert); with a maximum likelihood $f_p=\MLppstel$ ($\MLppplan$). Although we exclude the low-mass stellar companions, we do not exclude their parameter spaces from the completeness analysis. We do not expect companions of any mass near the known companions' semimajor axes; however, we compute posteriors in separation space, in which two companions may overlap. Further, we find that removing a vertical slice of the affected target stars' completeness maps does not significantly change their aggregated completeness.

Next, we look for correlations between planet and stellar properties and planet occurrence rate.

\subsection{Impact of planet separation}\label{ssec:separation}

\begin{figure*}[!htb]
    \centering
    \includegraphics[width=0.95\textwidth]{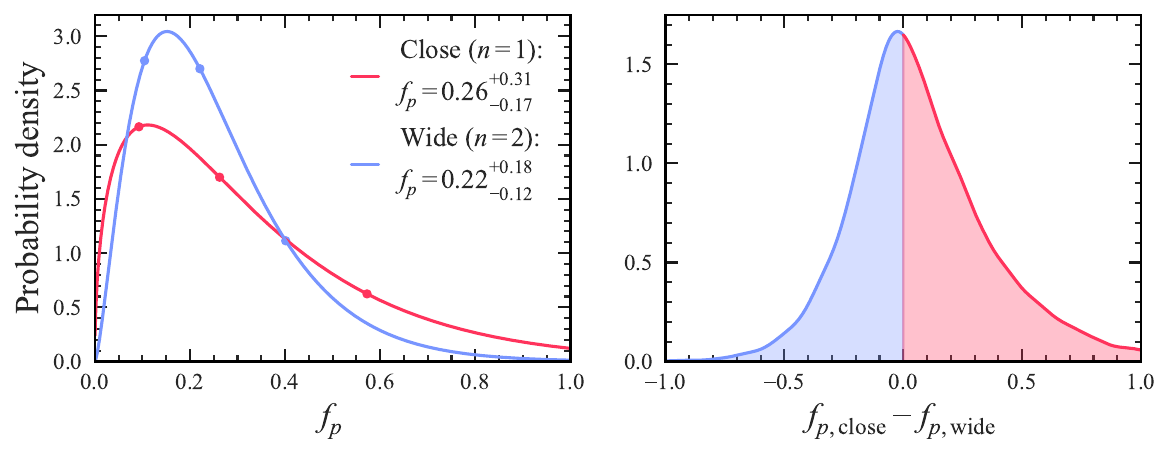}
    \caption{Left: posteriors on the occurrence rate of close ($<200$ mas, pink) vs. wide ($\geq 200$ mas, blue) companions, assuming the stellar accretion scaling. As in Figure~\ref{fig:posteriors_conf}, the legend provides the median and 16th and 84th percentiles, also marked by filled circles on the posteriors. Here we consider the three confirmed companions, one of which is ``close" (HD~142527~B) and two of which are ``wide" (PDS~70~c and HD~100453~B). Right: posterior on the \textit{difference} in planet occurrence rate at close vs. wide separations, $f_{p,\rm{close}}-f_{p,\rm{wide}}$; positive values (shaded in pink) indicate close companions are more common.}
    \label{fig:posteriors_closevswide}
\end{figure*}

Direct-imaging surveys for gas giants have demonstrated their low occurrence at wide ($\gtrsim 10$ au) separations, both by comparing the planet rate in bins of close and wide separation, and by inferring the power-law index of the semimajor axis distribution~(e.g.~\citealt{Nielsen2008,Wahhaj2013a,Kasper2007,Chauvin2010,Vigan2012,Vigan2017,Nielsen2019}). We set the boundary between ``close'' and ``wide'' to be 200~mas ($\approx 20$~au, for the average stellar distance of $\approx$100 pc). With this division, we compute the posteriors for a transitional disk system hosting close versus wide accreting companions. For a more conservative estimate, we again exclude the candidate companions LkCa~15~b and CS~Cha~c, but include the low-mass stellar companions. Of the remaining, HD~142527 B is ``close," and PDS~70 c and HD~100453 B are ``wide." The posteriors assuming the stellar accretion relation are shown in the left panel of Figure~\ref{fig:posteriors_closevswide}. The right panel shows the posterior on the \textit{difference} in occurrence rate, $\Delta f_p = f_{p,\rm{close}}-f_{p,\rm{wide}}$, found by drawing values from the individual posteriors and subtracting them. We find no evidence of a difference in accreting companion rate based on separation; while the inferred $f_p$ for close companions has longer tail, the uncertainties are so large that the posterior on $\Delta f_p$ is nearly centered on zero. This inconclusive result is unsurprising, given the small stellar sample size and few detections.

\subsection{Impact of stellar mass}

Searches for correlations between stellar mass and the planet occurrence rate have found the rate increases with mass, both in radial velocity surveys (e.g.~\citealt{Wolthoff2022}) and direct imaging (e.g.~\citealt{Bowler2016,Lannier2016,Galicher2016,Meshkat2017}). For the GPIES sample, \citet{Nielsen2019} both compare the rate in ``low-mass'' ($<1.5M_\odot$) and ``high-mass'' ($\geq1.5M_\odot$) bins and fit for a power-law dependence of $f_p$ with stellar mass. We adopt the same cutoff of $1.5M_\odot$ and compute the accreting companion occurrence rate around each subset of stars in the GAPlanetS sample. The masses for each survey star are given in Table~\ref{tab:surveystars}. In the GAPlanetS case, the two stellar companions orbit ``high-mass'' stars (8/14, or 57\% of the sample), while the lower-mass protoplanets and protoplanet candidates orbit ``low-mass'' stars (6/14, or 43\%, of the sample). This matches the intuition from selection effects. High-mass stars are intrinsically brighter: in the background-limited regime, the same achieved contrast means brighter, hence more massive, detectable objects. While we find the median inferred rate of accreting companions is slightly higher for the high-mass subsample, the rates for the low-mass and high-mass bins are consistent to well within uncertainty. The lack of evidence for a dependence of protoplanet frequency on stellar parameters is expected, given our sample size.

\section{Conclusion}\label{sec:conclusion}

Directly imaging protoplanets provides key insights into intermediate stages of planet and brown dwarf formation. With population inference, we can estimate the relative importance of different substellar formation pathways and the properties of the objects in each subpopulation. In this work, we present a method for quantifying survey completeness to accreting protoplanets, computing the first constraints on their population rate, and analyzing whether the rate depends on companion or stellar properties. The code for computing completeness is publicly available on GitHub.\footnote{\url{https://github.com/cailinplunkett/ProtoplanetPop}}

Determining accretion-rate and mass limits for protoplanets requires assuming astrophysical models that govern the conversions between these physical properties and an object's directly observable properties. We consider four main combinations of formation and accretion models in this work, though the technique is generalizable to any set of input relations. In particular, we compare two correlations between accreting objects' masses and mass accretion rates: one corresponding to a more ``star-like'', or isolated, process, estimated from empirical observations of low-mass stars and brown dwarfs (Equation~\ref{eq:stellarformation},~\citetalias{Betti2023}); and one pertaining to a more ``planet-like'' process, computed from simulations of circumstellar disk fragmentation (Equation~\ref{eq:planetformation},~\citetalias{Stamatellos2015}). For our constraints on protoplanet rate, we conservatively use a model that circumvents any assumption about the formation process, instead computing detectability as a function of \MMd. The accretion scalings describe the proportion of total accretion luminosity emitted in the observable \Ha line---a proxy for the physics of the accretion flows. We use one correlation derived from observations of accreting stars, corresponding to magnetospheric accretion (Equation~\ref{eq:stellaraccretion},~\citetalias{Alcala2017}), and one theoretical relation for a symmetric accretion shock on a planetary surface (Equation~\ref{eq:planetaccretion},~\citetalias{Aoyama2021}). We assume circumplanetary extinction is negligible.

We compute GAPlanetS' completeness to companions that formed and accreted under each combination of models. Using the formation-agnostic model, we infer the frequency of accreting companions from GAPlanetS for various subsets of the detected objects. The parameter space in which we infer the rate and the priors on the underlying \MMd and separation distributions have a strong effect on the inferred numbers. As such, we provide results for several possibilities for the parameter space and priors in Table~\ref{tab:rate_summaries}. We set a fiducial range of $-7\leq\log \MMd/(\MMdunits)\leq-3$, $30\leq\rm{sep \ [mas]}\leq2000$, and log-uniform priors on \MMd and separation. Under these assumptions, using the three confirmed accreting companions, the posterior median accreting companion occurrence rates are \mediancompratestell and \mediancomprateplan companions per star, assuming the stellar or planetary accretion mechanism, respectively. For objects within transitional disk gaps, the median rates are \medianingapstel (\medianingapplan) for the stellar (planetary) model. While one of these companions is a low-mass star, it differs from a field binary system, because it is embedded in a disk. Specializing to the single certain protoplanet, the median posterior rates are \medianppstel (\medianppplan). However, our small sample size contributes to broad CIs on the protoplanet occurrence rate. While we search for a dependence of the accreting companion rate on companion separation and stellar mass, our small number statistics and wide posteriors mean we do not find any correlation between companion rate and separation or stellar mass. Our current population analyses are suited to estimating rates under an assumed model, using subsets of the detections that we expect may have different properties. Currently, we do not have strong discriminating power on the population-level formation or accretion characteristics of protoplanets. Our inferred rates quantitatively reflect the difficulty of protoplanet imaging in \Ha.

Our results depend on several assumptions in the modeling process. Although we assume extinction local to the planet is small, if disk material tends to obscure accretion flows, fewer planets will be detectable in \Ha. Our completeness estimates must then be considered upper limits, and our companion rates conservative. As discussed in Section~\ref{ssec:formpar}, the planetary \MdtoM relation requires an assumption for the disk viscosity $\alpha$; lower viscosity disks produce higher accretion rates. A lower-viscosity assumption would result in higher completeness and lower inferred occurrence rates in turn. From Figure 8 in~\citetalias{Stamatellos2015}, the vertical offset of the \MdtoM scaling relation changes by $\lesssim$ one order of magnitude between $\alpha=0.001$ and $\alpha=0.05$, which would affect the search depth by at most the same factor. Although constraints on $\alpha$ for the GAPlanetS systems are weak, ideally one could adopt a distinct scaling relation for each target disk. As an extension of this work, the uncertainty in $\alpha$ could be marginalized over while simulating companions.

For tighter constraints on protoplanet rates, and to probe their population-level formation properties, we require a larger stellar sample size, lower uncertainties on confirmed companions' parameters, and more detections. The method presented here is generalizable to any combination of \Ha surveys. Incorporating additional survey stars necessitates the instrument parameters and stellar properties needed to compute contrast (Equation~\ref{eq:contrast}) as well as the contrast curve for each star. Computing the completeness and posteriors on accreting companion occurrence rate may then proceed as in Sections~\ref{sec:methods} and \ref{sec:constraints}. Planned \Ha surveys for protoplanets such as the MaxProtoPlanetS survey \citep{Close2020SPIE} will provide additional observations and, optimistically, detect new accreting objects. Alongside imaging more systems, both software and hardware developments will contribute to additional protoplanet detections.

An important avenue for increasing detections is improving the algorithms used to subtract the stellar PSF, thus probing higher contrasts and at tighter separations \citep[e.g.][]{Bonse2024}. Since planet frequency increases inward~\citep{Fernandes2019, Nielsen2019,Close2020}, improved reductions from such algorithms may lead to new protoplanet candidates in GAPlanetS (and other surveys') data. Instrument upgrades at Magellan \citep{Males2024} and the Very Large Telescope \citep{Boccaletti2022} will likely have similar effects, potentially revealing new protoplanet candidates in these systems by achieving higher contrasts at tighter inner working angles \citep{Close2020}. Updated constraints on the accreting companion rate can then be found by rerunning the algorithm presented here.

Incorporating results from other accretion-tracing emission lines that may be more prominent in planets, such as He~I or Pa$\beta$, is a promising avenue for detections, although it may complicate interpretations. For instance, He~I emission is also affected by winds~\citep{Thanathibodee2022,Erkal2022}. Refining the relations used to link emission-line accretion diagnostics to the underlying accretion rates of substellar is a critical avenue for upcoming theoretical and observational work (e.g.~\citealt{Follette2024jwst}).

Constraints on protoplanet population properties using protoplanet surveys in multiple tracers, uniformly processed using improved PSF subtraction algorithms, will be the subject of future work. By applying our framework to compute survey completeness to protoplanets, upcoming protoplanet surveys can be fully capitalized on, to bring us toward robust constraints on planet occurrence rates and formation mechanisms.

\section{Acknowledgments}

The authors thank the anonymous referee for the constructive review of this manuscript.  
KBF and CP acknowledge funding from NSF-AST-2009816.
G-DM acknowledges the support from the
European Research Council under the Horizon 2020 Framework Program via the ERC Advanced Grant ``Origins'' (PI: T. Henning), No.~832428, and via the research and innovation programme ``PROTOPLANETS'', grant agreement No.~101002188 (PI: M. Benisty), and
from the DFG priority program SPP 1992 ``Exploring the Diversity of Extrasolar Planets'' (MA~9185/1).

\appendix
\section{Statistical frameworks}\label{app:stats}

We infer the occurrence rate of protoplanets using a Bayesian approach, similar to the works of~\citet{Vigan2012}, \citet{Lannier2016}, \citet{Nielsen2019}, and \citet{Vigan2021}. In a survey of $N$ transitional disk stellar systems, for each star, we compute the detection probability as a function of \MMd and separation $s$, conditioned on astrophysical models $\Lambda$, $p_{\rm{det}}(\MMd,s\,|\,\Lambda)$. We denote with $f$ the true average number of accreting companions hosted by each system, with \MMd in the range $[\MMd_{\rm{min}},\MMd_{\rm{max}}]$ and separations $[s_{\rm{min}},s_{\rm{max}}]$. The fraction $f$ is assumed to be constant across the stellar sample; that is, we assume there is no correlation between stellar properties and companion rate.\footnote{One can relax this assumption by jointly inferring $f$ and its dependence on stellar properties.} We then compute the average completeness to objects in that parameter range, $p$, weighted by priors on $\MMd$ and separation:
\begin{align}
    p &= \frac1N\sum_{i=1}^N \frac{1}{(\MMd_{\rm{max}}-\MMd_{\rm{min}})(s_{\rm{max}}-s_{\rm{min}})}\int_{\MMd_{\rm{min}}}^{\MMd_{\rm{max}}} \int_{s_{\rm{min}}}^{s_{\rm{max}}} p_{\rm{det},i}(\MMd,s)\pi(\MMd)\pi(s)d(\MMd)ds\\
    &= \frac1N\sum_{i=1}^N p_i.
\end{align}

The assumed priors on separation and mass reflect assumptions about the underlying population. Observational results to date concur that planets at wide separations are rare and that lower-mass planets are more common. For our primary analyses, we use log-uniform priors on separation and \MMd: $\pi(s)\propto s^{-1}$, $\pi(\MMd)\propto \MMd^{-1}$. We provide rates for other prior choices in Table~\ref{tab:rate_summaries}. We have dropped the conditional on models $\Lambda$ for simplicity but this dependence is implicit in all the following. In practice, these integrals are approximated on a discrete grid using a trapezoidal sum.

We suppose that planet occurrence follows a Poisson process with rate parameter $f$. The expected number of detections is then $\langle n_{\mathrm{det}}\rangle = fpN$. The probability of observing $n$ detections given the rate $f$, the likelihood $\mathcal L(n|f,N,p)$, is
\begin{equation}\label{eq:likelihood_poisson}
    \mathcal L(n|f,p,N) = \frac{e^{fpN}(fpN)^n}{n!}.
\end{equation} An alternate approach is to use the Bernoulli likelihood, as in~\citet{Lannier2016,Squicciarini2025}. We find the Bernoulli likelihood gives nearly identical likelihoods to the Poisson, with marginally smaller uncertainties. Adopting a prior distribution on $f$, $\pi(f)$, Bayes' theorem states the posterior probability of rate $f$ given the data $n,N$ is
\begin{equation}\label{eq:posterior}
    P(f|n,p,N) = \frac{\mathcal{L}(n|f,p,N) \pi(f)}{\int \mathcal{L}(n|f,p,N) \pi(f)df},
\end{equation} where $P(n|p,N)= \int \mathcal{L}(n|f,p,N) \pi(f)df$ is the evidence.

\begin{figure}[!htb]
    \centering
    \includegraphics[width=0.48\textwidth]{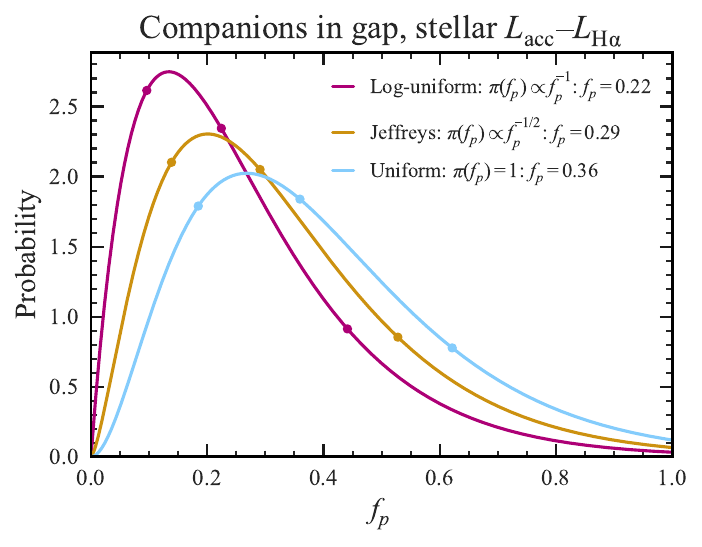}
    \caption{Comparison of posteriors on the companion rate under different prior assumptions: log-uniform (pink), Jeffreys (gold), and uniform (blue). These use the companions inside the gap and assume the stellar accretion scaling relation. The filled circles mark the median and 16th and 84th percentiles.}
    \label{fig:posteriors_comparepriors}
\end{figure}

The posteriors presented in Section~\ref{sec:constraints} are computed using a Jeffreys prior, $\pi(f) \propto f^{-1/2}$, for consistency with~\citet{Nielsen2019}. To test the importance of the assumed prior, we also compute posteriors using a log-uniform prior,  $\pi(f) \propto f^{-1}$, and a uniform prior $\pi(f)=1$. Since current results show planet rates are far below unity, the Jeffreys or log-uniform priors are most appropriate. Because of the small sample size and correspondingly broad likelihood distribution, the posteriors are somewhat prior-driven. As shown in Figure~\ref{fig:posteriors_comparepriors}, the log-uniform prior, which puts the most weight on low rates $f$, produces narrower posteriors with lower medians than the Jeffreys prior; using the uniform prior results in the widest posteriors with the highest inferred medians. The posteriors are nonetheless consistent to well within statistical uncertainty.

\bibliography{references}{}
\bibliographystyle{yahapj}
\end{document}